\documentclass[fleqn,usenatbib,useAMS]{mnras}

\usepackage{newtxtext,newtxmath}
 
\usepackage[T1]{fontenc}

\DeclareRobustCommand{\VAN}[3]{#2}
\let\VANthebibliography\thebibliography
\def\thebibliography{\DeclareRobustCommand{\VAN}[3]{##3}\VANthebibliography}

\usepackage{graphicx}	
\usepackage{amsmath}	
\usepackage{amssymb}	
\usepackage[usenames]{color}

\newcommand{\grad}{$^{\circ}$}

\title{The determination of asteroid $H$ and $G$ phase function parameters using GAIA DR2}

\author[M. Colazo et al.]{
M. Colazo,$^{1}$\thanks{E-mail: milagros.colazo@mi.unc.edu.ar}
R. Duffard,$^{3}$
W. Weidmann$^{2,4}$
\\
$^{1}$Instituto de Astronomía Teórica y Experimental (CONICET-UNC)\\
$^{2}$Observatorio Astronómico Córdoba, Universidad Nacional de Córdoba, Laprida 854, Córdoba, Argentina\\
$^{3}$Instituto de Astrofísica de Andalucía - CSIC. Granada, España\\
$^{4}$Consejo de Investigaciones Científicas y Técnicas de la República Argentina, Buenos Aires, Argentina
}

\date{Accepted XXX. Received YYY; in original form ZZZ}

\pubyear{2020}

\begin{document}
\label{firstpage}
\pagerange{\pageref{firstpage}--\pageref{lastpage}}
\maketitle

\begin{abstract}
The Gaia mission will provide the scientific community with high-quality observations of asteroids of all categories. The second release of Gaia data (DR2) was published in 2018 and consisted of 22 months of observations of 14,099 known Solar System objects, mainly asteroids. The purpose of this work is to obtain a catalogue of phase function parameters (H and G) for all the asteroids that are observed by the Gaia mission, and which were published in DR2. For this purpose, we introduced an algorithm capable of building this catalogue from the magnitude and UTC epoch data present in the DR2 database. Since Gaia will never observe asteroids with a phase angle of 0${^\circ}$ (corresponding with the opposition), but with phase angles higher than 10${^\circ}$, we added data from ground observations (corresponding to small phase angles) and thus improved the determination of the $H$ and $G$ parameters of the phase function. In this case, we also build a catalogue of the parameters of the H, G$_1$, G$_2$ phase function. We compared our results of the $H$, $G$ function with those of the Astorb database and observed that the level of agreement is satisfactory.
\end{abstract}

\begin{keywords}
minor planets, asteroids: general -- planets and satellites: fundamental parameters -- surveys
\end{keywords}



\section{Introduction}\label{Intro}

The Gaia mission will provide to the solar system scientific community with high-quality observations of about 300,000 asteroids of all categories. Most will be from the main belt but small asteroids from the NEA group, Jupiter trojans, centaurs and transneptunians are also observed. Gaia Data Release 2 (DR2)\footnote{\url{ https://www.cosmos.esa.int/web/gaia/data-release-2}} was published in 2018 and consisted of 22 months of observations of 14,099 known Solar System objects, mainly asteroids \citep{2010LNP...790..251H}, based on more than 1.5 million observations \citep{2018A&A...616A..13G}. As the database provided by Gaia, also another space and ground-based surveys are generating asteroid photometry data in large volume. As an example we can mention K2 \citep{2014PASP..126..398H, 2016PASP..128g5002V, 2016ksci.rept....6H}, TESS \citep{2015JATIS...1a4003R}, LSST \citep{Schwamb_2019, 2020arXiv200907653V} and VISTA \citep{2010iska.meetE...1A, 2017Msngr.167...16P}. K2 is the extension of the Kepler mission after one of its wheels failed. Since it points near the plane of the ecliptic, it can observe hundreds of asteroids \citep{2016AAS...22742102B}. K2 can measure the brightness of a large number of asteroids in the main belt almost continuously \citep{2016A&A...596A..40S}. TESS, like K2, can observe light curves nearly uninterrupted. In its case, it can do this for up to almost 27 days \citep{2018PASP..130k4503P}. The asteroids it observes are mostly those with inclinations greater than 6${^\circ}$ \citep{2020ApJS..247...26P}. An important difference between Gaia, K2 and Tess lies in the orbital position at which the asteroid is observed. In the case of TESS, the objects are observed during opposition, K2 observes them near their stationary points \citep{2018PASP..130k4503P}, and Gaia observes the asteroids when their phase angle exceeds 10${^\circ}$. The scientific community must be prepared to analyse this large amount of data that will allow us to encompass a more detailed knowledge of asteroid populations, which are currently poorly characterised, since basic physical properties such as mass, density, rotation properties, shape and albedo are not yet known for most of them.

We know that the brightness of asteroids varies both by their rotation on their axis and by their movement around the Sun. In this work we are interested in the second of these effects: the variation of the reduced magnitude (brightness on Johnson's filter $V$ normalised at 1 AU from the Sun and the observer, expressed in magnitudes) of the objects as a function of the phase angle $\alpha$.

This relationship is known as the phase function. In 1988 the IAU adopted the $H$, $G$ system to represent it: $H$ is the mean absolute magnitude in Johnson's V-band with a zero phase angle and $G$ is the "slope parameter" that describes the shape of the phase function. The absolute magnitude $H$ is defined as the brightness that the investigated object would have if it was located at 1 AU from both the Sun and the Earth and its phase angle was 0$^{\circ}$.

As the phase angle decreases, that is, as the asteroid approaches the opposition, the object will become brighter (its magnitude will decrease). For angles below 6$^{\circ}$ the decrease in magnitude is non-linear, which is called the \textit{opposition effect} \citep{2019P&SS..169...15C}. By contrast, for phase angles greater than 10$^{\circ}$ this relationship is practically linear. Gaia can obtain phase-magnitude curves precisely in this interval where the relationship between the $V$ and $\alpha$ parameters is linear \citep{2018A&A...616A..13G}. As \cite{2019P&SS..169...15C} mention, it is interesting to join the data of Gaia ($\alpha>10^{\circ}$) and data obtained from the ground ($\alpha<10^{\circ}$) to obtain determinations of $H$ of good quality that will serve, for example, to obtain estimations of the diameter of thousands of asteroids.

\noindent
Besides, \cite{2000Icar..147...94B}, proposed that the slope value of the linear part of the phase function can serve as an estimate of the albedo of the objects. Another interesting edge when analysing the Gaia data would be the properties of the phase-magnitude curves according to taxonomic type \citep{2019P&SS..169...15C}.

In a near future we need to be prepared to join together several and different asteroid data sources, obtained in different filters and from different sources/telescopes. We can reference all the data obtained to one reference standard like the Sloan filters or the Gaia photometric system.

The purpose of this work is to obtain a catalogue of phase function parameters (H and G) for all asteroids that are observed by the Gaia mission and published in DR2. This paper is organised as follows: In Section 2, we provide a brief description of the data used for the analysis for Gaia Data Release 2. In Section 3 we introduce the algorithm used to obtain the parameter catalogue. In Section 4 we discuss the results obtained and finally, in Section 5 we present the conclusions and future perspectives.

\section{Phase function equations}\label{phasefuneq}

\subsection{H, G magnitude phase function}\label{HG}

The $H$, $G$ phase function for asteroids can be described analytically through the following equation \citep{2010Icar..209..542M}:
\begin{equation}
    V(\alpha)=H-2.5\ \log_{10}[(1-G)\Phi_{1}(\alpha)+G\Phi_{2}(\alpha)], 
\label{eqn:2}
\end{equation}
where $\alpha$ is the phase angle, V($\alpha$) is the $V$ magnitude reduced to unit distance, $\Phi_{1}$($\alpha$) and $\Phi_{2}$($\alpha$) are two basis function normalised at unity for $\alpha$=0\grad. 

For the numerical calculation of the corresponding parameters, we follow the procedure indicated in \cite{2010Icar..209..542M}. The authors propose to write the equation \ref{eqn:2} as:
\begin{equation}
    10^{-0.4V(\alpha)}=a_{1}\Phi_{1}(\alpha)+a_{2}\Phi_{2}(\alpha),
\label{eqn:3}
\end{equation}
\noindent
where the absolute magnitude $H$ and the coefficient $G$ are:
\begin{align}
    H & =-2.5\ \log_{10}(a_{1}+a_{2}), &  G= & \frac{a_{2}}{a_{1}+a_{2}},    
\label{eqn:4}
\end{align}
\noindent
and the base functions can be accurately approximated by:
\begin{equation}
\begin{aligned}
    \Phi_{1}(\alpha) &= \exp & \left(-3.33\ \tan^{0.63}\frac{1}{2}\alpha \right ), \\
    \Phi_{2}(\alpha) &= \exp & \left(-1.87\ \tan^{1.22}\frac{1}{2}\alpha \right ).
\label{eqn:5}
\end{aligned}
\end{equation}

\noindent
Then, the coefficients a$_{1}$ and a$_{2}$ can be estimated using linear least squares. 

\subsection{H, \texorpdfstring{G$_1$}{G1}, \texorpdfstring{G$_2$}{G2} phase function}\label{HG1G2}

The three parameter magnitude phase function can be described as \citep{2010Icar..209..542M}:

\begin{equation}
    \begin{split}
    V(\alpha)  = H-2.5\log_{10}[G_{1}\Phi_{1}(\alpha) & +G_{2}\Phi_{2}(\alpha)\\
              & +(1-G_{1}-G_{2})\Phi_{3}(\alpha)],
    \label{eqn:6}
    \end{split}
\end{equation}

\noindent
where $\Phi_{1}$(0\grad)=$\Phi_{2}$(0\grad)=$\Phi_{3}$(0\grad)=1. Analogous to the previous model, the authors propose to write to the reduced magnitude $V$ as:
\begin{equation}
    10^{-0.4V(\alpha)}=a_{1}\Phi_{1}(\alpha)+a_{2}\Phi_{2}(\alpha)+a_{3}\Phi_{3}(\alpha).
\label{eqn:7}
\end{equation}

\noindent
The coefficients a$_{1}$,a$_{2}$ and a$_{3}$ can be calculated using minimum linear squares and then the parameters H, G$_{1}$ and G$_{2}$ are obtained with the following equations:

\begin{equation}
\begin{aligned}
    H &= -2.5 \log_{10}(a_{1}+a_{2}+a_{3}), \\
    G_{1} &= \frac{a_{1}}{a_{1}+a_{2}+a_{3}} ,\\
    G_{2} & = \frac{a_{2}}{a_{1}+a_{2}+a_{3}}.    
\label{eqn:8}
\end{aligned}
\end{equation}

\noindent
According to \cite{2010Icar..209..542M}, the $H$, $G$ function is reasonably good especially in the range of 10\grad to 60\grad. Also, it has the advantage of having a simple analytical form. However, there are cases where it produces bad adjustments even for high-quality observations. In contrast, the function H, $G_1$, G$_2$ does not present this problem. Having more parameters leads to better adjustments. This function, unlike $H$, $G$ can adequately adjust phase curves of asteroids with very high or very low albedo.

\section{Gaia Data}\label{gaia}
\begin{figure*}
    \centering
    \includegraphics[width=1\textwidth]{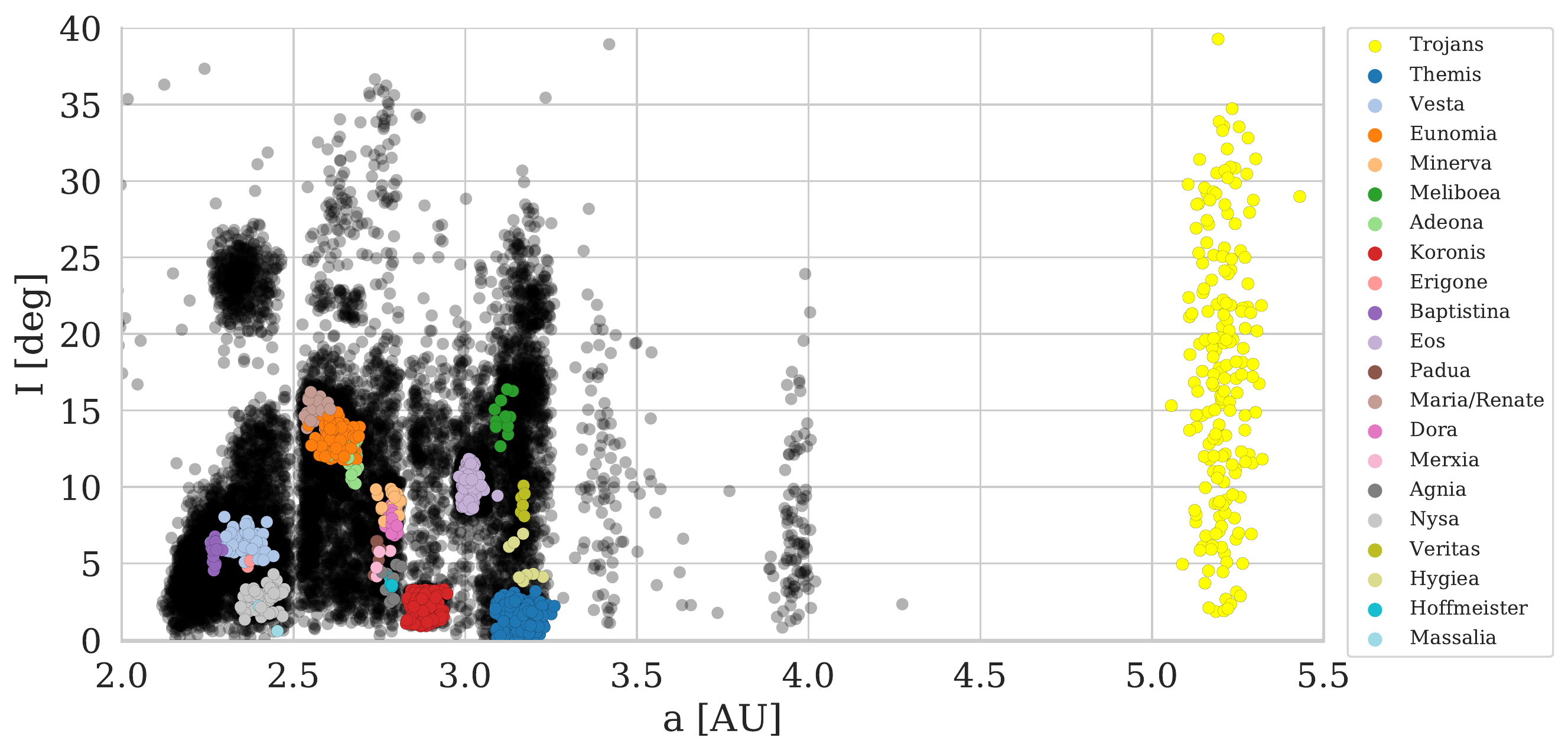}
    \caption{In black are represented all the
            asteroids observed by Gaia in the a (semimajor axis) vs I (inclination) plane. The families present in the sample and the Trojan asteroids are highlighted with different colours.}
    \label{fig:figure1}
\end{figure*}

As mentioned in Section \ref{Intro}, the Gaia DR2 was published in 2018 and had observations of 14 099 known objects from the solar system \citep{2010LNP...790..251H}. In this work, we have selected as sample the 13 981 objects that are asteroids. In Figure \ref{fig:figure1} we can see this sample represented in the $a$ (semimajor axis) vs $I$ (inclination) plane, where the different families and the Trojan asteroids are identified with different colours. 

We downloaded the database from the Gaia Archive\footnote{https://gea.esac.esa.int/archive/}. It has three columns: number mpc, epoch utc and $g$ magnitude (instrumental). The magnitude is published in the database and is obtained from the flux. We then also downloaded the flux and error flux to calculate the error in the magnitude by ourselves. In this work, we are interested in studying the variation of the magnitude according to the phase angle, therefore, we must process the data of the "raw" table. As a first instance, we need the phase angle and the reduced $g$ magnitude. The reduced magnitude is the one corrected to a 1 AU from the Sun and the Earth.

To obtain the phase angle, we start from the information of the epoch UTC. We must add a constant to this quantity to convert it to Julian date. The value of this constant is 2455197.5\footnote{\label{note1}\url{ea.esac.esa.int/archive/documentation/GDR2/}}. With the information of the Julian date, we perform a query to the Horizons ephemeris and save the information of the phase angle corresponding to the date.

\begin{figure}
    \centering
    \includegraphics[width=\columnwidth]{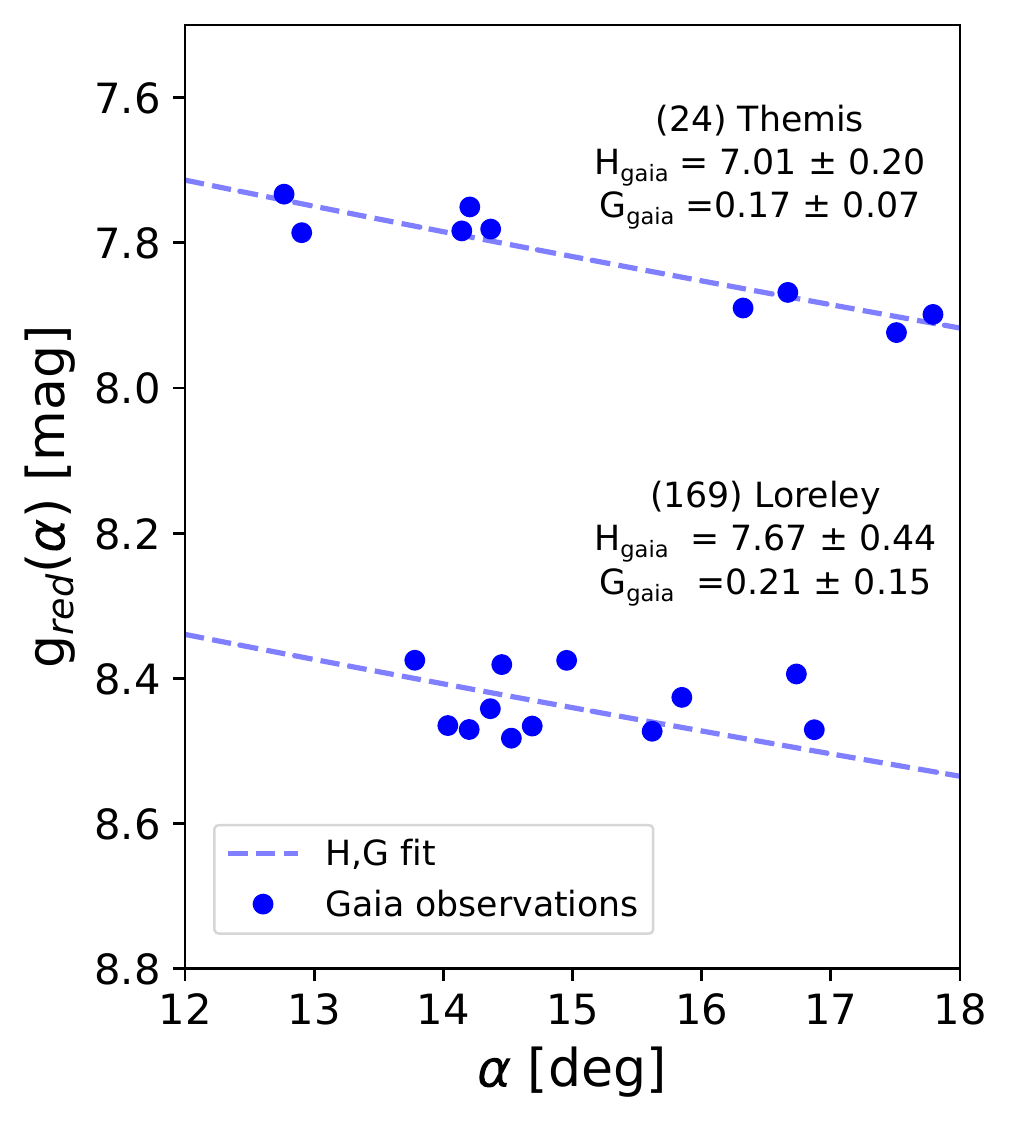}
    \caption{Phase curve of asteroid (24) Themis and (165) Loreley. In blue the observations of Gaia. In dashed line the adjustment of the $H$, $G$ system. The obtained H,G in these fits are in the $g$ band of Gaia.}
    \label{fig:figure2}
\end{figure}

To obtain the reduced $g$ magnitude, from the query to ephemeris we must store the information of r (gaiacentric distance) and $\Delta$ (heliocentric distance), both in AU. Then, we can calculate the reduced magnitude through the following equation:
\begin{equation}
    g_{\rm red} = g_{\rm inst} - 5\ \log_{10} (r.\Delta)
\label{eqn:9}
\end{equation}
\noindent
Once we have the phase angle and magnitude $g$ information reduced, we can adjust with the $H$, $G$ system (Equation \ref{eqn:2}) and plot the results. In Figures \ref{fig:figure2} we show the phase curves corresponding to asteroids (24) Themis and (165) Loreley, respectively. 

Paying attention in Figure \ref{fig:figure2} we can point up two problems presented by the Gaia data:

\begin{enumerate}
    \item \textbf{Dispersion by light curve}: We noted a large dispersion in the data obtained during a short time or in the same day or with similar phase angles. That could be due to the lightcurve of the asteroid, mainly the amplitude. In the case where there are well-sampled light curves observed from ground-based telescopes, usually select some criterion to choose the magnitude to be used for the calculation of phase functions. For example, \cite{2019P&SS..169...15C} always choose the maximum of their light curves. In the case of Gaia observations, this is not possible because there are no well-sampled light curves. Instead, what we have are a few observations of the asteroid per day. A priori, it is not possible to distinguish to which part of the asteroid's light curve those few observations correspond. This introduces a fairly wide dispersion in the phase curve, as shown in Figure \ref{fig:figure2}. For this work, to achieve some consistency given this issue, we have chosen the mean value of the magnitude in cases where there was more than one observation per day \citep{SHOWALTER2020114098}. 
    \item \textbf{Large phase angles}: Gaia can only observe asteroids that are at phase angles greater than 10${^\circ}$. Since we don't have information on small phase angles, the function settings are not as precise. Furthermore, to be able to adjust the H, G$_{1}$, G$_{2}$ function (Equation \ref{eqn:6}), this information is necessary. On the other hand, Gaia data do not allow us to model the opposition effect. To overcome this problem, we have supplemented Gaia data with observations taken from the ground. These observations do contain phase angles less than 10${^\circ}$.
\end{enumerate}

\subsection{Gaia data combined with ground-based observations}
The ground observations that were combined with the Gaia observations were taken from the Asteroid Photometric Catalog V1.0 \citep{1995PDSS....1.....L}. Five hundred asteroids from Gaia DR2 have been matched with the Asteroid Photometric Catalog V1.0. In this case, as we have complete light curves, we have chosen the minimum magnitude (maximum of the light curve) to carry out the phase function calculations. The magnitude data in this catalogue are in Johnson's filter V. In order to combine both data sets, we need to do the conversion from the reduced magnitude $g$ of Gaia to the reduced magnitude $V$ of Johnson. The equation that provides this connection between magnitudes is \citep{2018A&A...616A...4E}:

\begin{equation}
\begin{split}
    V = g_{\rm red} + 0.02269 - 0.01784 (V-R)  & + 1.016 (V-R)^2 \\
                                              & - 0.2225 (V-R)^3.  
\label{eqn:10}  
\end{split}
\end{equation}

\begin{figure}
    \centering
    \includegraphics[width=\columnwidth]{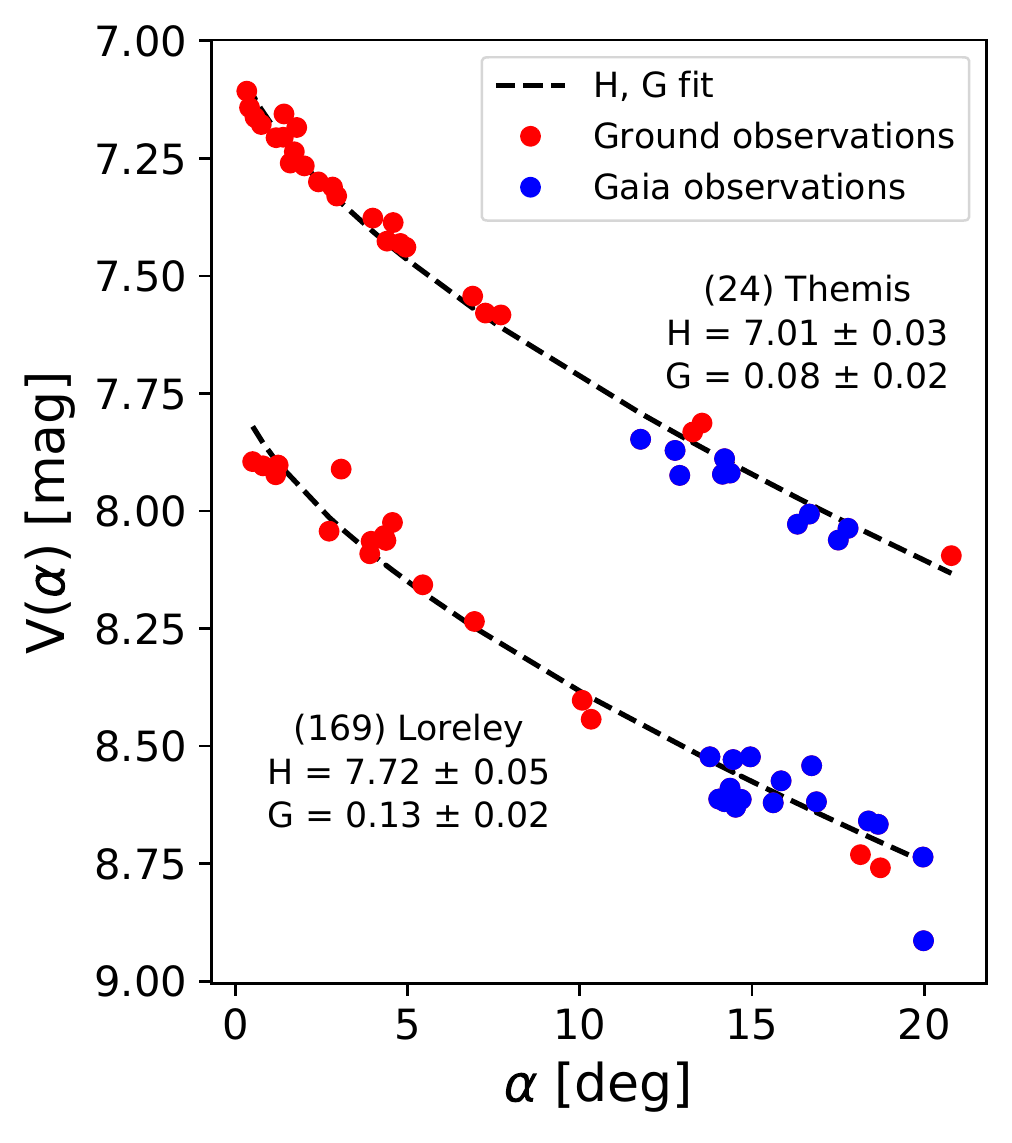}
    \caption{Phase curves of asteroids (24) Themis and (165) Leroley. In blue dots the observations of Gaia and in red dots, the ground observations. In dashed line the adjustment of the $H$, $G$ system.}
    \label{fig:figure3}
\end{figure}

\noindent
The standard deviation ($\sigma$) of the residuals of the fittings is 0.049 mag \citep{2018A&A...616A...4E}. This relationship is dependent on the colour of the object by its terms (V-R). In turn, the colours are related to the spectral type of the asteroid. \cite{2003Icar..163..363D} provide a table with measured (V-R) values for some spectral types. For those asteroids which have a taxonomic classification, we assign the (V-R) of the table from \cite{2003Icar..163..363D} for the calculation of the $V$ magnitude. However, some problems arise again:
\begin{enumerate}
    \item There are many more spectral classifications than those shown in the table. In the case that the spectral classification of the asteroid was not in the mentioned table, we assigned the value of (V-R) corresponding to the taxonomic type with which it shares more similarities.
    \item There are few asteroids that have spectral classification. In cases where the asteroid has no spectral classification, we have assigned a $(V-R)_{\rm average} = 0.43$ that was obtained by averaging all values in the table. This is the mean for all the spectral types. A more precise way to carry out this calculation would be to obtain an average colour index weighted according to the abundance of the asteroids population because the most numerous asteroids are C and S type. However, C-type asteroids have $V-R = 0.37$ and  S-type asteroids have V-R = 0.47. If we use a mean value between these 2 spectral types, we should use $V-R = 0.42$ which is very close to the $V-R = 0.43$ that we used.

\end{enumerate}
\noindent
Once the magnitudes $g$ of Gaia have been passed to Jhonson's magnitude V, we repeat the calculation of the system $H$, $G$ and we plot the phase curve. Again, to facilitate the comparison with Figure \ref{fig:figure2}, we show the phase curves corresponding to (24) Themis and (165) Leroley in Figure \ref{fig:figure3}.
\noindent
The fact that the phase angle range is sampled more completely is evident. Let us also note that the error in the calculation of the parameters decreases significantly.

\begin{figure}
    \centering
    \includegraphics[width=\columnwidth]{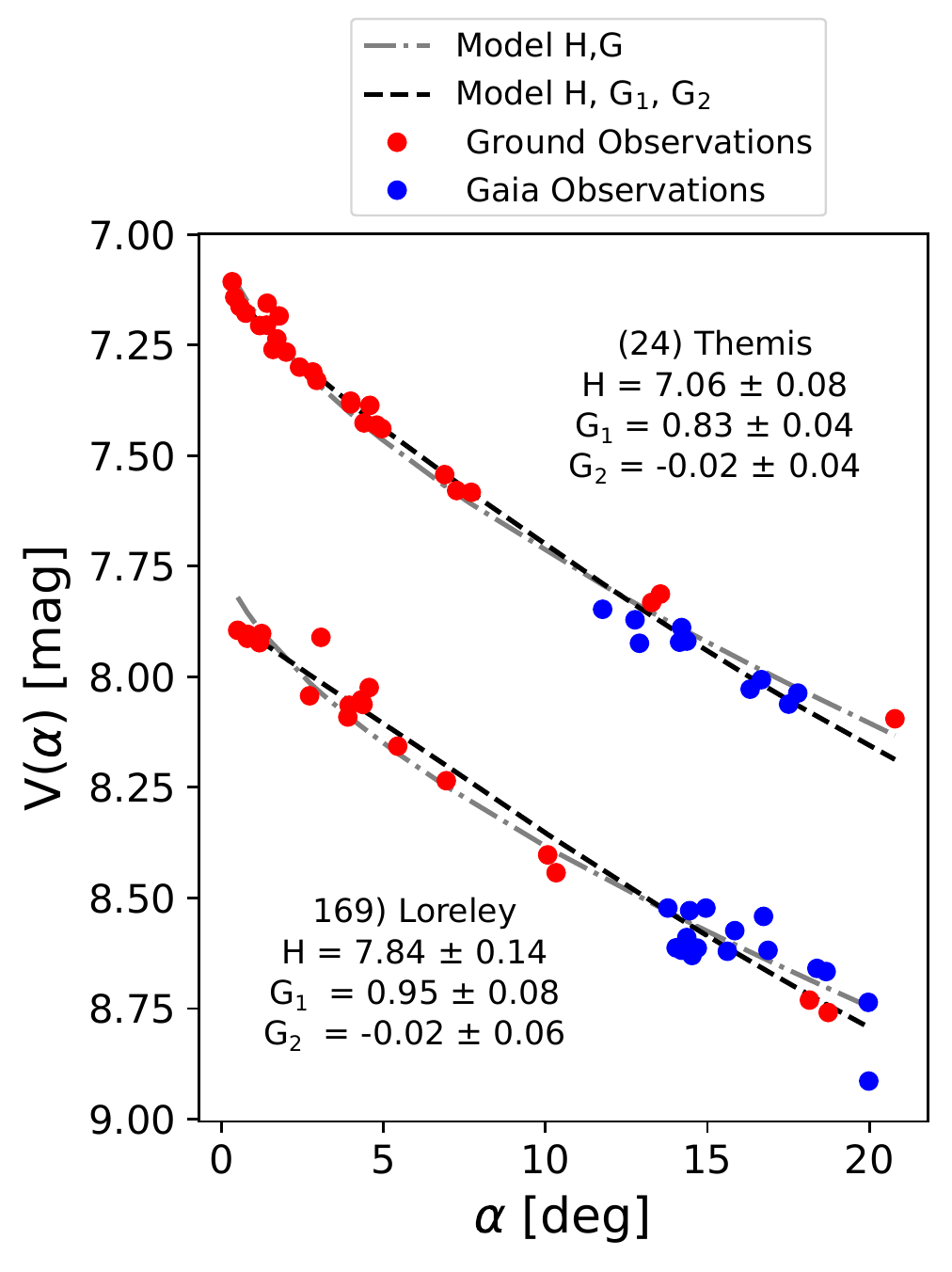}
    \caption{Phase curves of asteroids (24) Themis and (165) Leroley. In blue dots the observations of Gaia and in red dots, the ground observations. In black dashed line the adjustment of the H, G$_{1}$, G$_{2}$ system. For comparison, in grey dashed line the adjustment of $H$, $G$ function.}
    \label{fig:figure4}
\end{figure}

A further step that we can achieve when we are working with Gaia data combined with ground data is to fit the function H, G$_{1}$, G$_{2}$. In Figure \ref{fig:figure4} we can see the adjustment of this function for the same two asteroids shown above. For comparison, we have also included the fit with the $H$, $G$ function. 

In this case, we can see that the error in the determination of $H$ with Equation \ref{eqn:6} is somewhat greater than with the Equation \ref{eqn:2}. This behaviour may be due to the fact that the dispersion of Gaia data affects this model more than the other. The reason is that according to \cite{2016P&SS..123..117P}, the function H, G$_1$, G$_2$ can present problems when applied to observations of low quality and/or with large variations in magnitude.

\section{Results}
In this Section we introduce the results obtained after processing the data with our implementation in Python. We will divide this section according to the set of data used for the fit. This will clearly show the differences between using only Gaia data and its magnitude $g$ and using Gaia data in magnitude $V$ combined with ground-based observations.
In all the cases where the adjustment has been made with the $H$, $G$ model presented in Section \ref{HG}, the value obtained for the parameter $H$ was compared with the one present in Astorb's database\footnote{\url{https://asteroid.lowell.edu/main/astorb}}.

We have filtered our final catalogues retaining only those asteroids whose error in the $H$ magnitude is less than 30$\%$ of the adjusted $H$ value. This criterion was used in all three cases: a) $H$, $G$ with only Gaia data in magnitude $g$ (9,817 asteroids), b) $H$, $G$ with Gaia data in magnitude $V$ combined with ground observations (480 asteroids) and c) $H$, G$_{1}$, G$_{2}$ with Gaia data in magnitude $V$ combined with ground observations (190 asteroids).

\subsection{Brief discussion about errors}
For this brief error analysis, we have selected the same objects that we have been using as examples in the previous Sections: (24) Themis and (165) Leroley.
The errors of $H$ presented in this paper correspond to the errors given by the least-squares adjustments. The objective of this brief analysis is to show that the errors in the magnitudes $g$ and $V$ are negligible compared to the error produced by the fits.
The error in Gaia magnitude $g$ can be estimated from the flux data present in DR2. According to the documentation \citep{2018gdr2.reptE...5B}:

\begin{equation}
    \begin{aligned}
              g = & -2.5 \log(g_{\rm flux}) + G_{0}, \\
            G_{0} &= 25.6884 \pm 0.0018.
    \end{aligned}
    \label{eqn:11}
\end{equation}

\noindent
To obtain the error in magnitude $g$ we propagate errors.
On the other hand, to estimate the error in $V$ magnitude all we do is to make error propagation in Equation \ref{eqn:10} assuming that the error in (V-R) is 0.1.

In the case of (24) Themis, the errors obtained for the magnitudes are $error_g = 0.005$ and $error_V = 0.009$. The errors in the determination of the $H$ parameter for the different fits are: 0.2 ($H$, $G$ phase function using only Gaia data), 0.03 ($H$, $G$ phase function combining Gaia data and Earth observations) and 0.08 ($H$ phase function, G$_1$, G$_2$ combining Gaia data and Earth observations).

On the other hand, for (165) Leroley, we have obtained the following results: $error_g = 0.007$ and $error_V = 0.01$. In this case, the errors in the determination of the parameter $H$ for the different fits are: 0.44 ($H$, $G$ phase function using only Gaia data), 0.05 ($H$, $G$ phase function combining Gaia data and Earth observations) and 0.14 ($H$ phase function, G$_1$, G$_2$ combining Gaia data and Earth observations).

As we can see, the errors of the parameter $H$ are always greater than the errors estimated for the magnitudes $g$ and V.

\subsection{Gaia data in g magnitude.}\label{GG}
In this subsection, we show the data obtained by using only the data from Gaia observations in their magnitude g. To perform the adjustment, it will only be necessary to identify the phase angle corresponding to the observation and calculate the reduced magnitude g. It is important to note that in this case, it is not necessary to know the spectral type information (colour V-R) of the asteroids.

\begin{figure}
    \centering
    \includegraphics[width=\columnwidth]{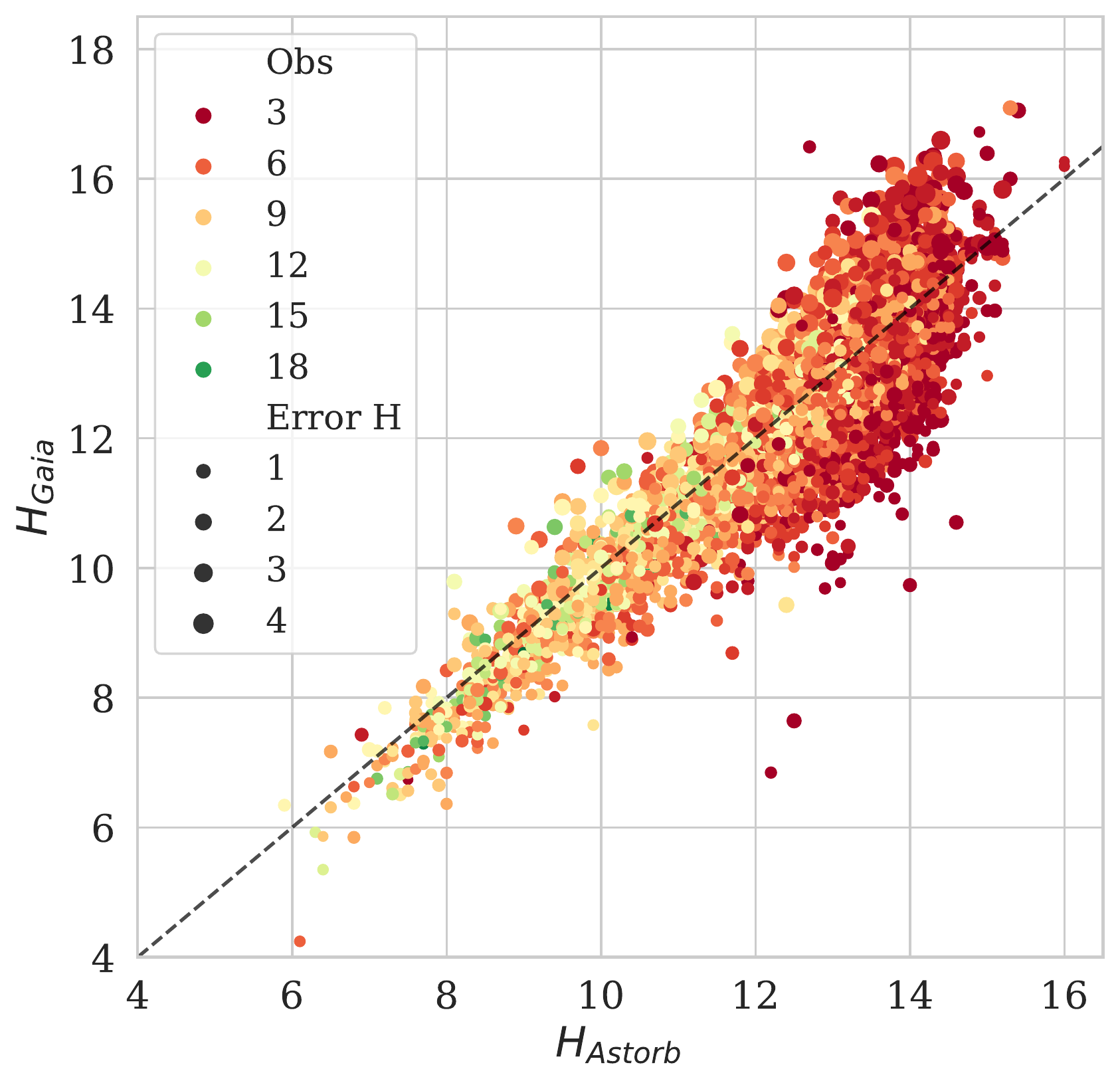}
    \caption{Comparison of the obtained $H$ using Gaia data in $g$ magnitude with the one
    published in the Astorb database. The colours represent the number of observations, and the point size represents the error. The dashed line indicates the line of coincidence. The figure illustrates 9,817 asteroids.}
    \label{fig:figure5}
\end{figure}

In Figure \ref{fig:figure5} we present the comparison of our results of the parameter $H$ with those published in Astorb. The dashed line indicates the line of coincidence. The number of observations is represented in colour (green for the most observed and red for the least observed) and the point size represents the error (the larger the point size, the greater the error). We can notice that those asteroids that have fewer observations are also those that have larger errors. Besides, this group belongs to those objects that are fainter.

\begin{figure}
    \centering
    \includegraphics[width=\columnwidth]{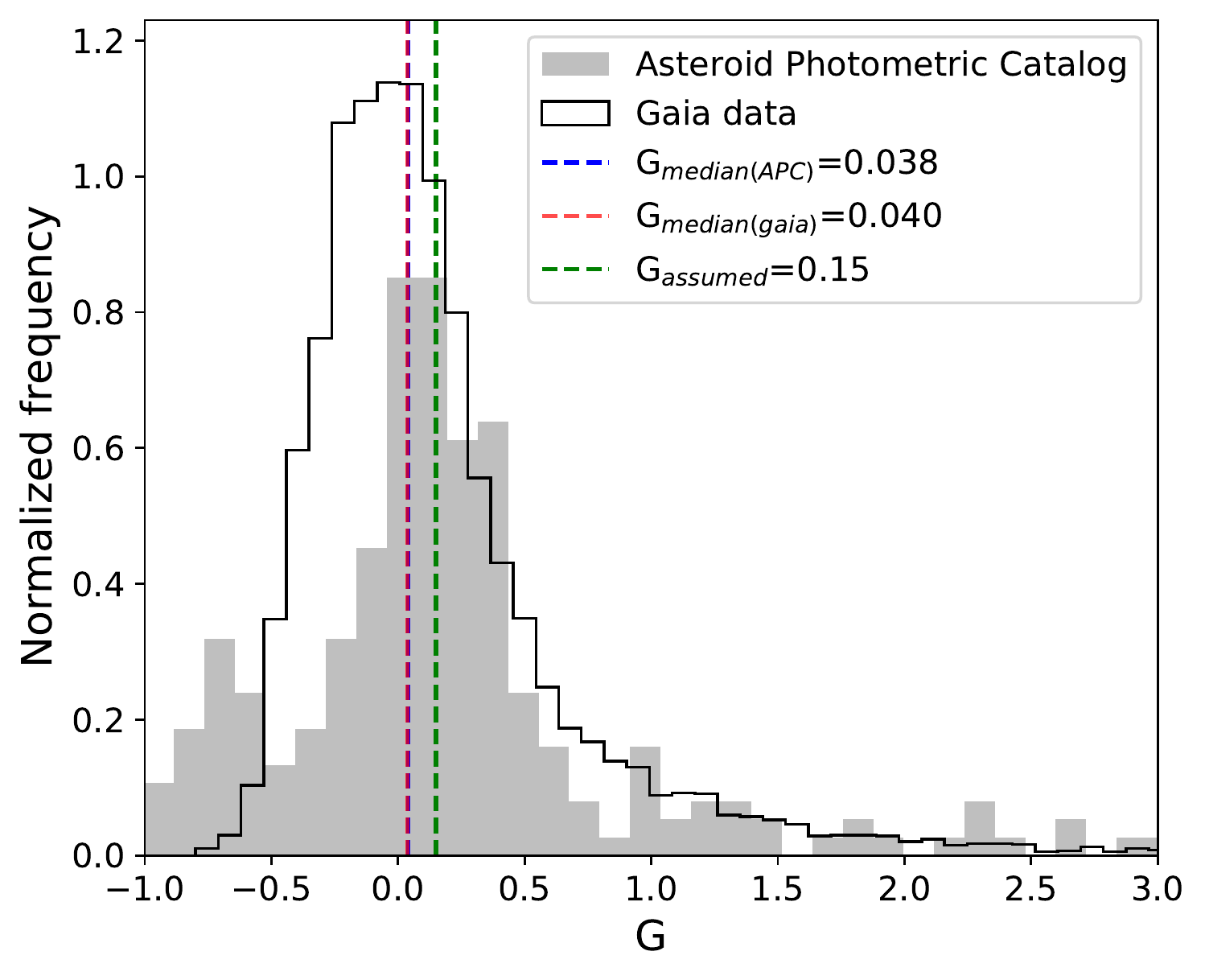}
    \caption{Histogram of G-values obtained using Gaia (9,817 asteroids). The green dotted line marks the value $G = 0.15$ that is assigned to most asteroids. The red line denotes the sample median. The grey histogram corresponds to the distribution of $G$ from the Asteroid Photometric Catalogue (APC), 313 asteroids. In blue dotted line the median of this sample. Note that the Gaia median and the APC median are superposed.}
    \label{fig:figure6}
\end{figure}

\noindent
We also obtained the $G$ parameters for those asteroids. The values obtained using the Gaia data are shown in
the histogram in Figure \ref{fig:figure6}. It is important to note that we have calculated $G$ parameters that have a wide range of values as opposed to the fixed value $G = 0.15$ that is assigned to most of them as established in MPC 17257 (December 1990). The median of our sample is 0.04. We have included in the same figure the distribution of $G$ for the Asteroid Photometric Catalogue, to have an immediate comparison between the two sources. The median, in this case, is also 0.04.

\begin{figure}
    \centering
    \includegraphics[width=\columnwidth]{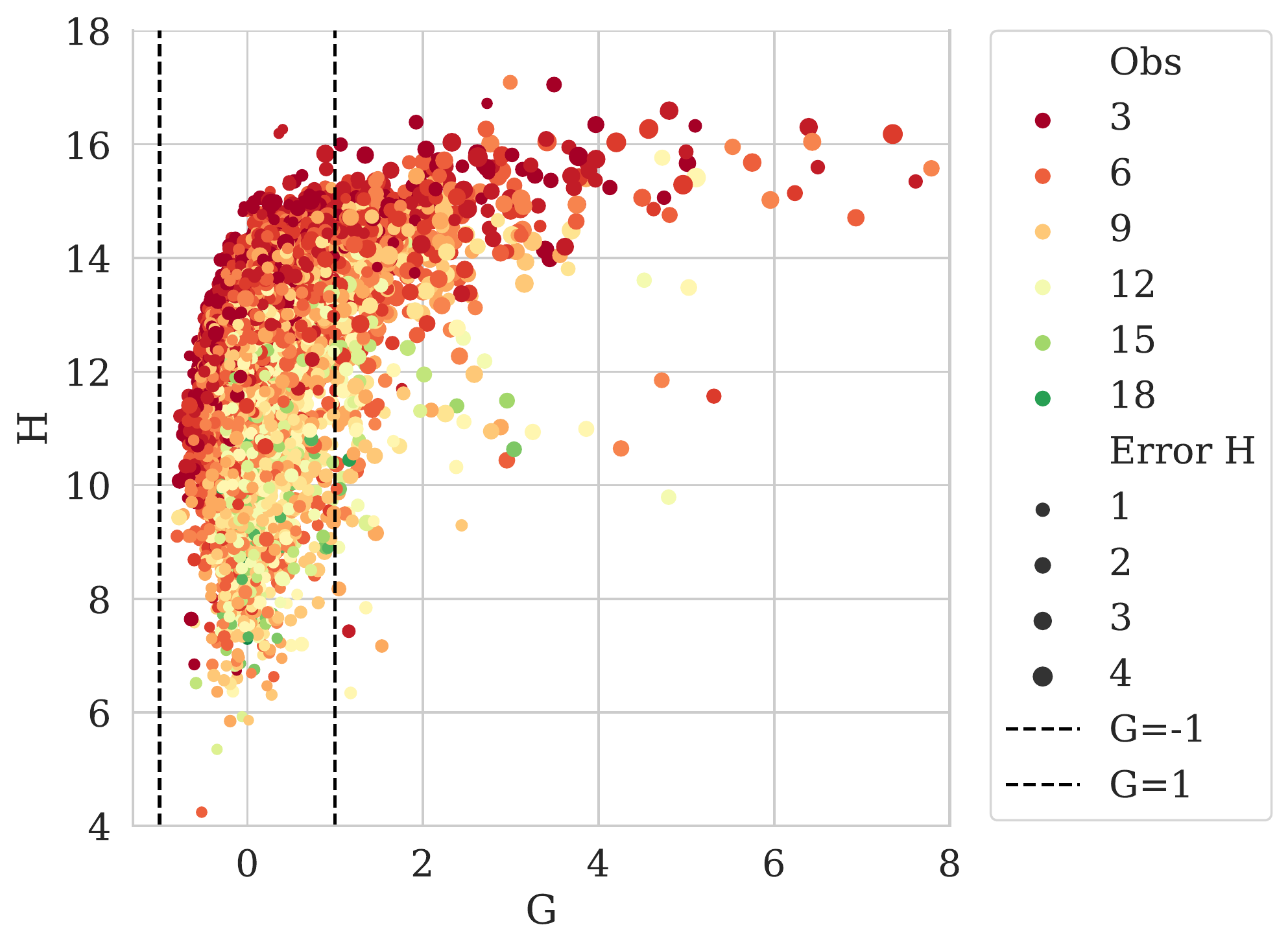}
    \caption{Graph $G$ vs $H$ for Gaia data in $g$ magnitude. The colour of the points represents the number of observations.}
    \label{fig:figure7}
\end{figure}

\noindent
Another interesting correlation that can be analysed is that of $G$ vs $H$. Looking at Figure \ref{fig:figure7} we notice that the asteroids whose parameter $G$ is greater than one are those that have few observations and that also have relatively high absolute magnitudes. Those that are brighter and have a
considerable number of observations (green dots) tend to be
placed on a vertical line around $G = 0$. It is to be expected that
as more observations are obtained, the red dots that form the sort of "hump" will flatten out and become more vertical.

\subsection{Gaia data combined with ground observations.}
Here we present the same plots as in the previous subsection but transforming Gaia $g$ magnitude into Jhonson's $V$ magnitude. At this point, it becomes essential to know the spectral information of the asteroids, since it is needed to carry out the passage between magnitudes. We will see how the determination of $H$, $G$ parameters improves when including data from ground observations. The number of points decreases dramatically because the sample size of the Asteroid Photometric Catalog V.1 is much smaller than the Gaia sample. 

On the other hand, since we now have information for <10\grad angles because we include ground observations, we also calculate the fit with the H, G$_{1}$, G$_{2}$ model presented in Section \ref{HG1G2}.

\begin{figure}
    \centering
    \includegraphics[width=\columnwidth]{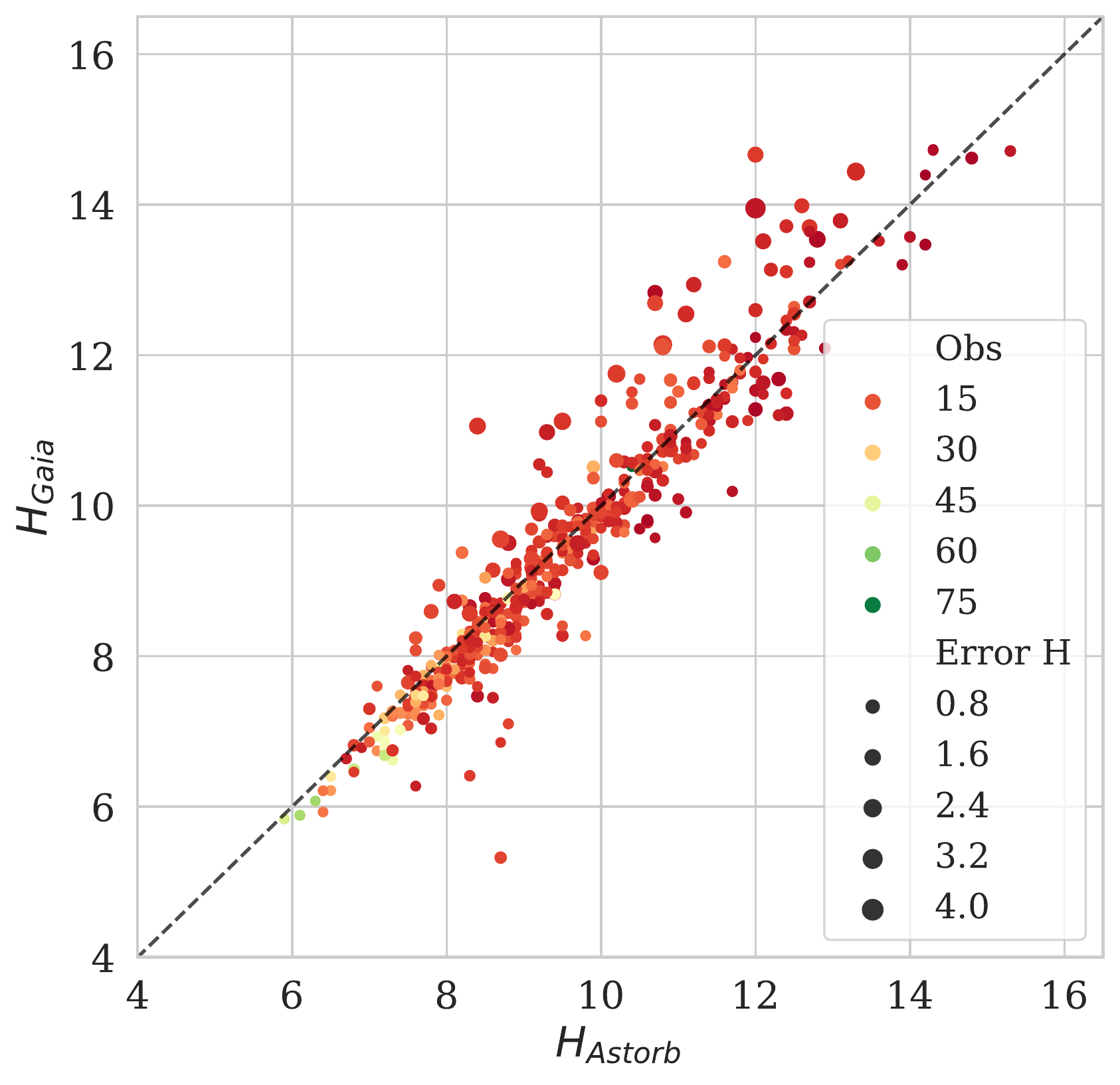}
    \caption{Comparison of the obtained $H$ using Gaia data in $V$ magnitude combined with ground observations, with the one published in the Astorb database. The colours represent the number of observations, and the point size represents the error. The dashed line indicates the line of coincidence.}
    \label{fig:figure8}
\end{figure}

In Figure \ref{fig:figure8} (analogue to the Figure \ref{fig:figure5}), $H$ parameters were calculated by adding observations from the ground (small phase angles). It is important to note the decrease in the dispersion after combining the Gaia mission data with the photometric data obtained from the ground. This is because the Gaia mission does not observe phase angles smaller than 10\grad, which are observed from the ground. When combining both sets of observations, a more complete scan of the phase angles is obtained. 

\begin{figure}
    \centering
    \includegraphics[width=\columnwidth]{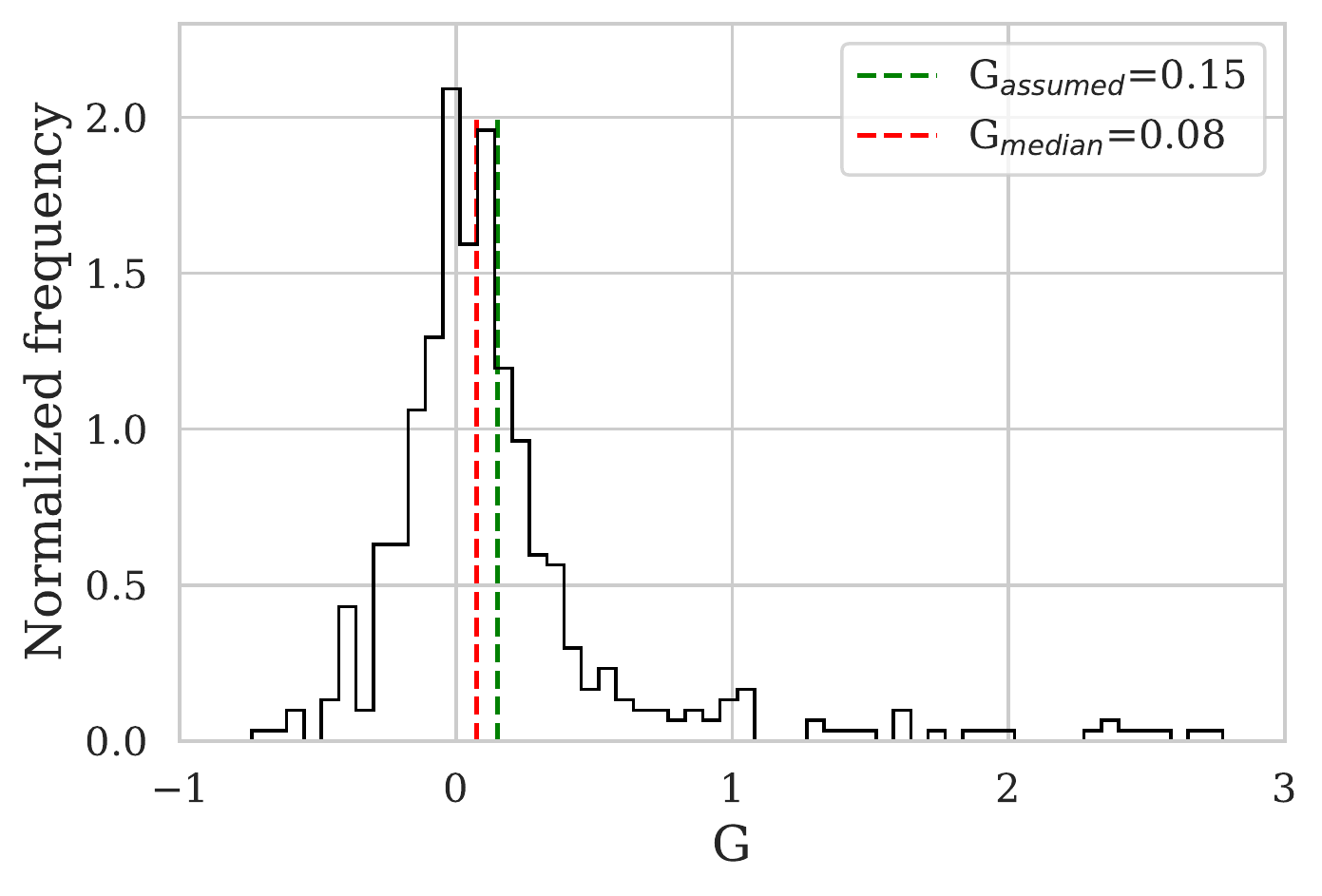}
    \caption{Histogram of $G$ values obtained using Gaia data and data from ground-based observations. The green dotted line marks the value $G = 0.15$ that is assigned to most asteroids. The red line denotes the sample median.}
    \label{fig:figure9}
\end{figure}

\noindent
We enhanced the determination of $G$ parameters by adding observations from the ground where available. Analogous to Figure \ref{fig:figure6} we present the histogram of the obtained values in Figure \ref{fig:figure9}. We denote the sample median which in this case is $G = 0.08$ and the value $G = 0.15$ which is the one generally assigned to the asteroids.

\begin{figure}
    \centering
    \includegraphics[width=\columnwidth]{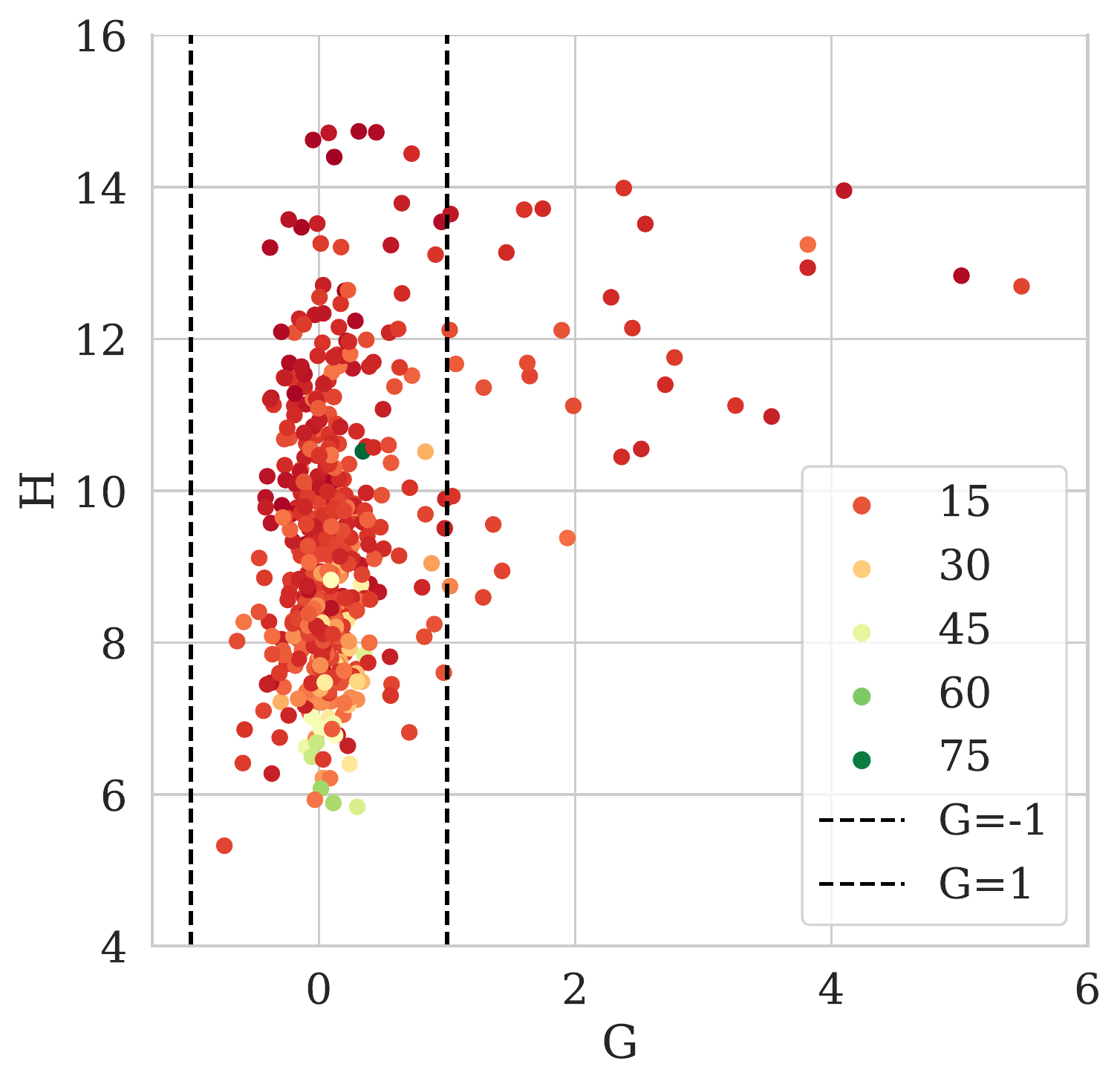}
    \caption{Graph $G$ vs $H$. The colour of the points represents the number of observations.}
    \label{fig:figure10}
\end{figure}

\noindent
In Figure \ref{fig:figure10} we show the $G$ vs $H$ plot. In this case, where we combine Gaia data with ground-based observations, we see that most points are located vertically around $G = 0$.

In Section \ref{GG} $H$ and $G$ are computed with Gaia data, i.e. with large phase angle (> 10\grad ), so it is reasonable to expect values of $G$ close to zero because the opposition effect is completely excluded. In this section, the parameters $H$ and $G$ are computed using Gaia data plus the phase curves for smaller angles obtained from APC. In this case, we expect more realistic results for $H$ and G. For this reason, we compare them with high-quality data \citep{2012Icar..221..365P} and with another large database from Pan-STARRS \citep{2015Icar..261...34V}. 

\cite{2012Icar..221..365P} provide accurate $H$ and $G$ slope parameters for more than 580 asteroids with densely covered light curves in a single pass band over a wide range of phase angles and sets the standard in measuring asteroid photometric properties. In Pravec’s main-belt sample the median $G$ value is 0.23, while the median of our sample is considerably lower: 0.08. Also the Pravec’s $H$ vs. $G$ plot (i.e. the equivalent of Figure \ref{fig:figure10}) is not centered on zero and appears much less dispersed as regards the values of G.

On the other hand, \cite{2015Icar..261...34V} show the distribution for 250,000 asteroids with a peak near $G\approx0.15$ (Figure 9) i.e., near the value usually employed. Again, the median value of our sample is significantly below the one calculated by these authors.

To compare the three samples, we reproduce in Table \ref{tab: 1}, the obtained values in Table 5 from \cite{2015Icar..261...34V}. In this table, one can compare the obtained mean values of $G$ for different taxonomic classes. In this case, when discriminating by taxonomic class, we note that the values obtained in this work agree reasonably well with those published by the other authors. Actually, for S and C types our values are closer to those of \cite{2012Icar..221..365P} than those obtained by \cite{2015Icar..261...34V}. On the other hand, we note that our data have higher values of standard deviations.

So, while the median $G$ may be underestimated when looking at the full Gaia + APC sample, when separating by taxonomic type we see that the values obtained are within the expected range considering high quality observations such as those of \cite{2012Icar..221..365P}.

\begin{table}
\caption{Mean slope parameters $\pm$ standard deviation derived in this work (Gaia $+$ APC, first column), by \protect\cite{2015Icar..261...34V} (PS1, second column) and by \protect\cite{2012Icar..221..365P} (PRA12, third column) for the same objects in three major taxonomic classes. In the fourth column we indicate the number of asteroids in the sample corresponding to the taxonomic type for Gaia $+$ APC. For taxonomic types Q and D we were unable to make a comparison. The Gaia $+$ APC sample does not possess asteroids of taxonomic type Q.}
\label{tab: 1}
\begin{tabular}{ccccc}
\hline
\multicolumn{1}{|c|}{\begin{tabular}[c]{@{}c@{}}Taxonomic \\ Class\end{tabular}} &
\multicolumn{1}{c|}{Gaia + APC} &
  \multicolumn{1}{c|}{PS1} &
  \multicolumn{1}{c|}{PRA12} &
  \multicolumn{1}{c|}{N} \\ \hline
Q & n/a & $0.11\pm0.16$ & $0.19\pm0.10$ & 0 \\
S & $0.19\pm0.44$ &  $0.16\pm0.26$ &  $0.23\pm0.05$ & 139\\
C & $0.13\pm0.63$ & $0.03\pm0.10$ & $0.13\pm0.01$ &  107\\
D & $1.1\pm2.9$ & \protect{n/a} & \protect{n/a} & 14 \\
X &  $0.16\pm0.51$ & $0.21\pm0.30$   & $0.20\pm0.10$  & 69 \\
\end{tabular}
\end{table}

\begin{figure}
    \centering
    \includegraphics[width=\columnwidth]{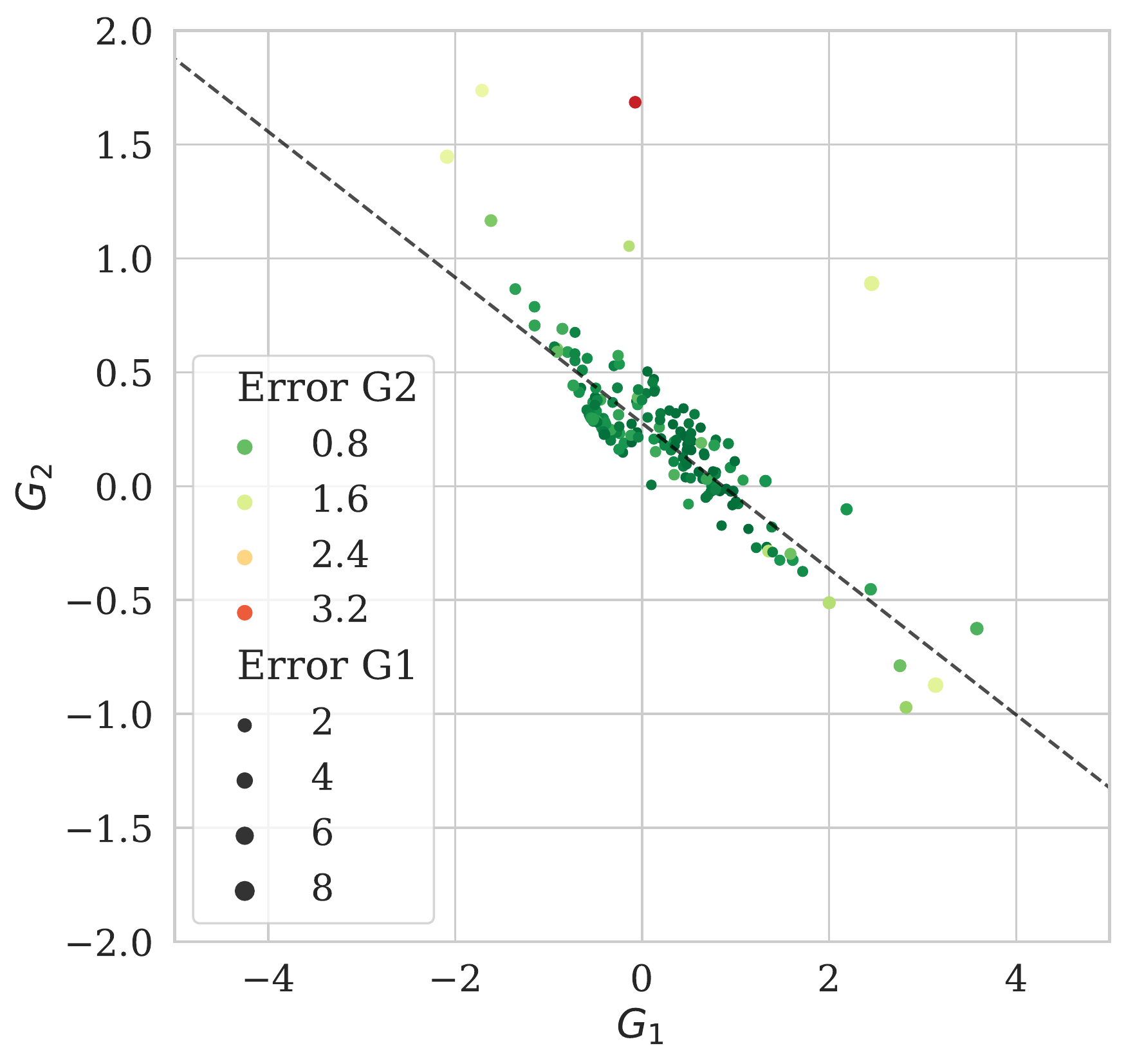}
    \caption{Graph $G$ vs $H$. The colour of the points represents the number of observations.}
    \label{fig:figure11}
\end{figure}

\noindent
Finally, for the fit of the H, G$_{1}$, G$_{2}$ model, we show in Figure \ref{fig:figure11} a plot of G$_{1}$ vs G$_{2}$ that was published previously in other works (\cite{2010Icar..209..542M, 2019P&SS..169...15C}). Our results, as well as those obtained by the previous authors, show an approximately linear correlation between both parameters.

\section{Conclusions}
For the final conclusions of our work, we believe it is useful to analyse the strengths and weaknesses of the Gaia DR2 data.

Using only Gaia data in magnitude $g$ can be useful as a first approximation of the parameters, but they are not very reliable. The fact that these observations do not cover small phase angles and that they also present a notorious dispersion introduces errors in the determination of the parameters. 

The attractive work arises from combining Gaia data with ground-based observations, as the determination of parameters is considerably improved. This is because by using both data sets, we obtain a more complete sampling in the phase angle range. However, some drawbacks arise: 1) there are few asteroids with light curves for a wide range of phase angles. This becomes a major problem when combining these terrestrial data with a database as extensive as the Gaia Data Release. 2) In some cases, even though the asteroid has a good number of observations from the Earth, the scatter introduced by the Gaia data "disrupts" the adjustment.  3) A few asteroids have spectral classifications (V-R colour information). This introduces errors in passing Gaia magnitude $g$ to $V$ to combine with ground data. By conducting analyses such as those presented in this paper, valid results can be obtained for "statistical" physical studies, placing our obtained values in the level (2) of accuracy proposed by \citep{2010Icar..209..542M}.

As future work, it is interesting to explore the correlation of the parameters G, G1 and G2 with the albedo of the asteroids and the influence that these can have to make the spectral classification. On the other hand, with our extensive catalogue of absolute magnitudes $H$ we can estimate diameters for thousands of objects and compare our results with the NEOWISE database \citep{2014ApJ...792...30M}. 

Another attractive possibility in the future of this work is to combine different asteroid databases, as already proposed by other authors \citep{2018A&A...617A..57D, 2020A&A...643A..59D}. We can join data from space (Gaia, K2, Tess, WISE \citep{2010AJ....140.1868W}) with data obtained from Earth (VISTA, Pan-STARRS \citep{2018AAS...23110201C}, Zwicky \citep{2014htu..conf...27B}, Tomo-e \citep{2016cosp...41E2051W, 2018SPIE10702E..0JS}, Sloan \citep{2006AJ....131.2332G}, own observations). As we mentioned in Section \ref{Intro}, the data of Gaia, K2 and TESS complement each other since they observe the asteroid in different positions of its orbit. In this way, we will include both sparse data and densely sampled light curves, exploiting to the fullest extent all the information we have at our disposal. 

\section*{Acknowledgements}
We thank the anonymous reviewer for his valuable comment on our manuscript, which have certainly raised its quality. Funding from Spanish project AYA2017-89637-R is acknowledge. Financial support from the State Agency for Research of the Spanish MCIU through the Center of Excellence Severo Ochoa for the Instituto de Astrofisica de Andalucia (SEV-2017-0709). Milagros Colazo is a doctoral fellow of CONICET (Argentina).  

\section*{Data Availability}
The data underlying this article were accessed from the Gaia
archive, \url{https://gea.esac.esa.int/archive/}. The derived
data generated in this research will be shared on reasonable request
to the corresponding author.



\bibliographystyle{mnras}
\bibliography{mnras} 




\bsp	
\label{lastpage}
\end{document}